\documentclass{article}

\PassOptionsToPackage{numbers, compress}{natbib}
\usepackage[final]{nips_2018}

\usepackage[utf8]{inputenc} 
\usepackage[T1]{fontenc}    
\usepackage{hyperref}       
\usepackage{url}            
\usepackage{booktabs}       
\usepackage{amsmath}
\usepackage{mathtools}
\usepackage{amsfonts}       
\usepackage{nicefrac}       
\usepackage{microtype}      
\usepackage{mycommands}
\usepackage{xspace}
\usepackage{authblk}
\usepackage{dsfont}
\usepackage{graphicx}
\usepackage{multirow}
\usepackage{siunitx}
\usepackage[symbol]{footmisc}

\graphicspath{{figures/}} 

\title{Efficient prediction of 3D electron densities using machine learning}

\begin{document}

\author[1]{Mihail Bogojeski\thanks{Contributed equally.}}
\author[1,2]{Felix Brockherde$^*$}
\author[3]{Leslie Vogt-Maranto$^*$}
\author[4]{Li Li\thanks{Currently at Google AI}}
\author[3,5,6]{Mark E. Tuckerman}
\author[4,7]{Kieron Burke}
\author[1,8,9]{Klaus-Robert Müller}

\affil[1]{\footnotesize Machine Learning Group, Technische Universität Berlin, Marchstr. 23, 10587 Berlin, Germany}
\affil[2]{\footnotesize Max-Planck-Institut für Mikrostrukturphysik, Weinberg 2, 06120 Halle, Germany}
\affil[3]{\footnotesize Department of Chemistry, New York University, New York, NY 10003, USA}
\affil[4]{\footnotesize Departments of Physics and Astronomy, University of California, Irvine, CA 92697, USA}
\affil[5]{\footnotesize Courant Institute of Mathematical Science, New York University, New York, NY 10003, USA}
\affil[6]{\footnotesize NYU-ECNU Center for Computational Chemistry at NYU Shanghai, 3663 Zhongshan Road North, Shanghai 200062, China}
\affil[7]{\footnotesize Department of Chemistry, University of California, Irvine, CA 92697, USA}
\affil[8]{\footnotesize Department of Brain and Cognitive Engineering, Korea University, Anam-dong, Seongbuk-gu, Seoul 02841, Korea}
\affil[9]{\footnotesize Max-Planck-Institut für Informatik, Stuhlsatzenhausweg, 66123 Saarbrücken, Germany}

\maketitle

\begin{abstract}
The Kohn–Sham scheme of density functional theory is one of the most widely used
methods to solve electronic structure problems for a vast variety of atomistic
systems across different scientific fields. While the method is fast relative
to other first principles methods and widely successful, the computational time needed is still not negligible,
making it difficult to perform calculations for very large systems or over
long time-scales. In this submission, we revisit a machine learning model capable
of learning the electron density and the corresponding energy functional based on
a set of training examples. It allows us to bypass solving the Kohn-Sham
equations, providing a significant decrease in computation time. We specifically
focus on the machine learning formulation of the Hohenberg-Kohn map and its decomposability.
We give results and discuss challenges, limits and future directions. 
\end{abstract}

\section{Introduction}
The electron density $n(\vr)$ is one of the fundamental properties of atomistic systems.
According to the first Hohenberg-Kohn theorem of density functional theory (DFT) the electron
density uniquely determines the ground state properties of an atomistic system~\cite{hohenberg1964inhomogeneous}.
Kohn-Sham density functional theory~\cite{kohn1965self} (KS-DFT) provides a relatively efficient framework for calculating the
electronic energy, making it one of the most popular electronic structure methods across a wide array of fields~\cite{pribram2015dft}.

Recently, there has been an increase in the application of machine learning (ML) methods
to various problems regarding atomistic systems\cite{rupp2018special}. Such machine learning models have been applied
for the prediction of properties of molecules and materials by learning from a large database of
reference calculations~\cite{rupp2012fast,de2016comparing,jose2012construction,hansen2013assessment,
schutt2017quantum,faber2017prediction,schutt2017schnet,pronobis2018many,yao2018tensormol},
performing molecular dynamics (MD) by learning the potential energy surfaces/force fields for
particular molecules~\cite{chmiela2017machine,li2015molecular,grisafi2018symmetry,chmiela2018towards,
gastegger2017machine, glielmo2018efficient,zhang2018deep,christensen2018operators}, and in few cases for the
prediction of electron densities as a means of performing electronic structure calculations
in the DFT framework without solving the expensive Kohn-Sham equations~\cite{snyder2012finding,li2016pure,
brockherde2017bypassing,seino2018semi,welborn2018transferability,anton2018deep,grisafi2018transferable}.
These machine learning methods are often able to predict properties of atomistic systems
or perform MD simulations at a similar accuracy to DFT calculations while requiring a
fraction of the computation costs, which could potentially allow computational scientists to examine larger systems
and/or longer time-scales.

In this submission we will revisit a paper by Brockherde et al.~\citep{brockherde2017bypassing}, which deals
with bypassing the Kohn-Sham equations by first learning the Hohenberg-Kohn (HK) map
from the one-body potential $v(\vr)$ to the electron density $n(\vr)$ of an atomistic
system. Subsequently a second model is used to learn the energy functional
from the predicted density to the total energy of the system. This way, 
the model provides both approximations to the electron density as well as accurate potential energy surfaces
suitable for molecular dynamics simulations of small atomistic systems.

Here we will focus on the machine learning approach used, especially for the Hohenberg-Kohn map, 
and explain in a more didactic manner how this approach enables us to predict the density using
multiple independent models, making the learning problem much simpler.

\begin{figure*}
  \centering
  \includegraphics[width=.8\textwidth]{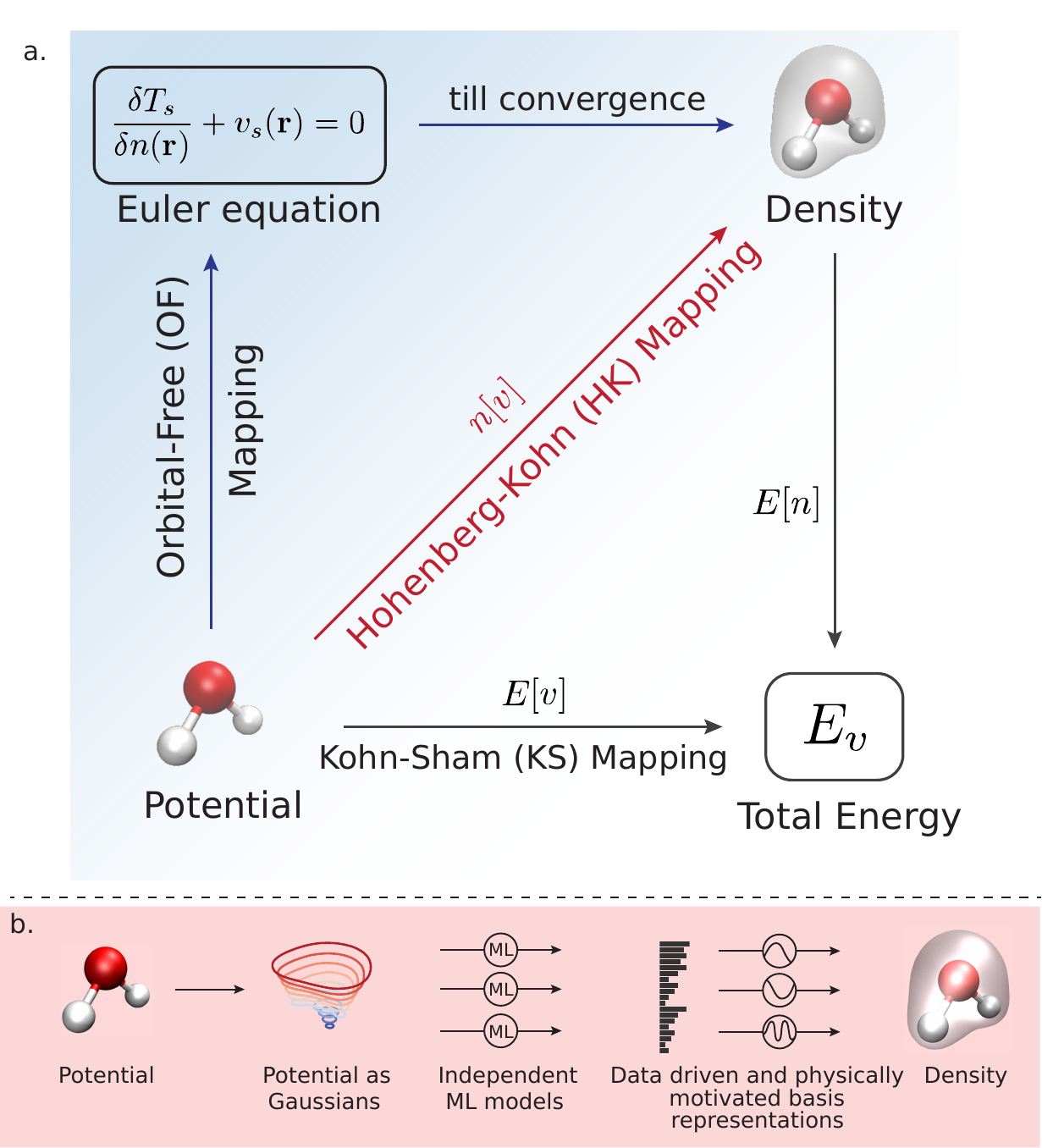}
  \caption{\label{fig:model} \textbf{Overview of the ML-HK map}
    \textbf{a.} General overview of machine learning models for the total energy. The bottom arrow represents $E[v]$, 
    a conventional electronic structure calculation, i.e., KS-DFT\@. The ground state energy is found by solving KS 
    equations given the external potential, $v$. $E[n]$ is the total energy density functional. The red arrow is 
    the HK map $n[v]$ from external potential to its ground state density.
    \textbf{b}. How the machine learning Hohenberg-Kohn (ML-HK) map makes predictions.
    The molecular geometry is represented by Gaussians; many independent Kernel Ridge Regression models predict each
    basis coefficient of the density. We analyze the performance of data-driven (ML) and common physical basis
    representations for the electron density. Figure adapted from Brockherde et al.~\cite{brockherde2017bypassing}.
  }
\end{figure*}

\section{Methods}
\subsection{Kohn-Sham density functional theory (KS-DFT)}\label{sec:ks-dft}
The KS-DFT computational electronic structure method is based on the Hohenberg-Kohn theorem~\cite{hohenberg1964inhomogeneous}
which establishes the unique correspondence between the potential and electron density, i.e.~at most one potential
can produce a given ground-state electron density. KS-DFT can thus be used to calculate various properties
of many-body atomistic systems using functionals of the electron density.

The KS-DFT scheme models a fictitious system of non-interacting electrons whose
density is the same as the real one, thereby avoiding direct calculation of
the many-body effects~\cite{kohn1965self}. The accuracy of KS-DFT is limited by the accuracy of the
used approximation to the unknown exchange-correlation energy, whereas the main computational bottleneck
is computing the solution of the KS equations describing the non-interacting orbitals, which has a
complexity of $\mathcal{O}(n^3)$.

All 3D DFT calculations used in this submission are performed  with the Quantum ESPRESSO 
software~\cite{giannozzi2009quantum}, using the PBE exchange-correlation functional~\cite{perdew1996generalized}
and projector augmented waves (PAWs)~\cite{kresse1999ultrasoft,blochl1994projector} with Troullier-Martin
pseudization for describing the ionic cores~\cite{troullier1991efficient}.
All molecules are simulated in a cubic box (L~=~20~bohr) with a wave function cutoff of 90~Ry.

\subsection{Kernel ridge regression (KRR)}
KRR~\cite{hastie2009elements} is a machine learning method for non-linear regression. Non-linearity is
achieved by incorporating 
the kernel trick into linear ridge regression, which finds the optimal linear mapping from the
inpus to the labels under $\ell_2$ regularization. Let $\x_1, \dots, \x_M \in \R^d$ be the training data points and let
$\Y = {\left[\y_1, \dots, \y_M \right]^T}$ be their respective labels. The KRR model for a
new input sample $\x^*$ is then given by:
\begin{align}
  \y^* = \sum\limits_{i=1}^M \alpha_j k(\x^*, \x_i),
\end{align}
where $k$ is a kernel function and $\balpha = [\alpha_1, \dots, \alpha_M]^T$ are the model weights.
The model weights are obtained by solving the following optimization problem:
\begin{align}\label{eq:krr_min}
  \min\limits_{\balpha} \left\{ \sum\limits_{i=1}^m \left|\y_i - \sum\limits_{j=1}^m \alpha_j k(\x_i, \x_j)\right|^2 + \lambda \balpha \K\balpha \right\}
\end{align}
where $\lambda$ is a regularization parameter and $\K$ is the kernel matrix
with $\K_{ij} = k(\x_i, \x_j)$. The analytical solution to the minimization problem is then given by
\begin{align}\label{eq:krr_sol}
    \balpha = {\left(\K + \lambda \I\right)}^{-1} \Y.
\end{align}
Here we use the Gaussian (radial basis function) kernel
\begin{align}
    k(\x, \x') = \exp\left(-\frac{||\x - \x'||^2}{2\sigma^2}\right),
\end{align}
where the kernel width $\sigma$ is a model parameter that needs to be tuned using cross-validation.

\subsection{Machine learning Hohenberg-Kohn (ML-HK) map}
The ML-HK map was introduced in a paper by Brockherde et al.~\citep{brockherde2017bypassing} as a way to avoid minimizing
the total energy using gradient descent, requiring the calculation of gradient the kinetic energy
model, which is often unstable due to missing information about direction outside of the data manifold~\cite{snyder2013kernels}.

As an alternative, the ML-HK map is a multivariate machine learning model that directly learns the electron density $n(\vr)$
as a functional of the potential $v(\vr)$, thereby completely bypassing the need for calculating the kinetic energy gradient
(see Figure~\ref{fig:model}a).

As a first step, we model the potential of a system using an artificial Gaussians potential, which is calculated as:
\begin{align}
    v(\vr) = \sum\limits_{\alpha=1}^{N^\text{a}} Z_\alpha \exp \left( \frac{-\lVert \vr - \Rr_\alpha \rVert^2}{2\gamma^2} \right),
\end{align}
where $\Rr_\alpha$ is the position and $Z_\alpha$ the nuclear charge of the $\alpha$-th atom. The resulting
artificial potential is evaluated on a 3D grid around the molecule and is used in this form as a descriptor for the ML
model. Cross-validation can be used to optimize the width $\gamma$ of the artificial potential as well
as the spacing $\Delta$ of the grid. We will use $\vv$ to denote vector representation of the potential evaluated 
on a 3D grid.

Using this artificial potential as the descriptor, a naive formulation of the ML-HK map would be
\begin{align}
  n^\text{ML}[\vv](\vr) = \sum_{i=1}^M \beta_{i}(\vr) k(\vv, \vv_i).
\end{align}
In this formulation, each grid point $\vr$ has its own set of model weights $\bbeta(\vr)$, which means that we
would need to train a separate model for each grid point of the density. Given the cubic growth of the 
potential and density grids, it is easy to see that this approach would quickly become intractable even for small molecules. 
Additionally, nearby grid-points are strongly correlated, and using independent models for nearby points would essentially
disregard the information contained in the local correlations.

To circumvent this obvious drawback, we use a basis representation for the density of the form
\begin{align}
  n(\vr) = \sum_{l=1}^L u^{(l)}\phi_l(\vr),
\end{align}
where $\phi_l$ are basis functions and $u^{(l)}$ are the corresponding basis coefficients. Here we use the Fourier
basis representation for the electron density, however many other basis representations such as kernel
PCA\cite{brockherde2017bypassing} or atom-centered Gaussian type orbitals \cite{grisafi2018transferable} can also be used.
With this formulation, we can transform the learning problem from one of predicting the density grid points to one 
of predicting the basis coefficients $u^{(l)}[v]$, giving us the following model for the predicted density
\begin{align}
  n^\text{ML}[\vv](\vr) = \sum_{l=1}^L u^{(l)}[\vv]\phi_l(\vr).
\end{align}
Using KRR, the model for each coefficient can be formulated as
\begin{align}
  u^{\text{ML}(l)}[\vv] = \sum_{i=1}^M \beta_i^{(l)} k(\vv, \vv_i),
\end{align}
with $\bbeta^{(l)}$ representing the model weights associated with each basis coefficient and $k$ being the Gaussian kernel.
The resulting contributions of the prediction error to the cost function are given by
\begin{align}
  \begin{split}
    err(\bbeta) &= \sum\limits_{i=1}^M \lVert n_i - n^\text{ML}[\vv_i] \rVert_{\mathcal{L}_2}^2\\
            &= \sum_{i=1}^M \left\lVert n_i - \sum_{l=1}^L u^{\text{ML}(l)}[\vv_i]\phi_l \right\rVert_{\mathcal{L}_2}^2.
  \end{split}
\end{align}
By writing the density in terms of its basis representation and assuming orthogonality of the basis functions we obtain
\begin{align}
  \begin{split}
    err(\bbeta) &= \sum_{i=1}^M \left\lVert \sum_{l=1}^L u_i^{(l)}\phi_l - \sum_{l=1}^L u^{\text{ML}(l)}[\vv_i] \phi_l \right\rVert_{\mathcal{L}_2}^2\\
            &= \sum_{i=1}^M \left\lVert \sum_{l=1}^L\left( u_i^{(l)} - u^{\text{ML}(l)}[\vv_i] \right)\phi_l  \right\rVert_{\mathcal{L}_2}^2\\
            &= \sum_{i=1}^M \int \sum_{l=1}^L\left( u_i^{(l)} - u^{\text{ML}(l)}[\vv_i] \right)\phi_l(\vr)
            \sum_{l'=1}^L\left( u_i^{(l')} - u^{\text{ML}(l')}[\vv_i] \right)\phi^*_{l'}(\vr) d\vr\\
            &= \sum_{i=1}^M\sum_{l,l'=1}^L\left(u_i^{(l)} - u^{\text{ML}(l)}[\vv_i] \right)
            \left(u_i^{(l')} - u^{\text{ML}(l')}[\vv_i] \right)\int \phi_l(\vr)\phi^*_{l'}(\vr)d\vr\\
            &= \sum_{i=1}^M\sum_{l=1}^L\left(u_i^{(l)} - u^{\text{ML}(l)}[\vv_i] \right)^2\\
            &= \sum_{i=1}^M\sum_{l=1}^L\left(u_i^{(l)} - \sum_{j=1}^M \beta_j^{(l)} k(\vv_i, \vv_j) \right)^2.
  \end{split}
\end{align}

The resulting equation shows that the error can be decomposed into the independent error contributions for each of the 
basis coefficients. By viewing the errors independently we obtain $L$ separate KRR minimization problems, and analogously to
equations~\ref{eq:krr_min} and~\ref{eq:krr_sol} we obtain the analytical solutions
\begin{align}
\bbeta^{(l)} = {\left(\K_{\sigma^{(l)}} + \lambda^{(l)} \I\right)}^{-1} \vu^{(l)}, \quad l = 1, \dots, L,
\end{align}
where for each basis function $\phi_l$, $\lambda^{(l)}$ is a regularization parameter, $\vu^(l)$ is a vector containing
the training set coefficients for the $l$-th basis function and $\K_{\sigma^{(l)}}$ is a Gaussian kernel matrix with width $\sigma^{(l)}$.

By independently and directly predicting the basis coefficients, the machine learning model becomes more efficient and easier to
scale to larger molecules. Additionally, the basis representation allows us to use the predicted coefficients to
reconstruct the continuous density at any point in space, making the predicted density independent of a fixed
grid and enabling computations such as numerical integrals to be performed on the predicted density at an arbitrary accuracy.

Finally after the ML-HK map is learned, modelling the energy functional requires a single KRR model of the form
\begin{align}
  E^{ML}[n] = \sum\limits_{i=1}^M \balpha_j k(\vu^{\text{ML}}[\vv], \vu^{\text{ML}}[\vv_i]).
\end{align}
where $\vu^{\text{ML}}[\vv] = \left[u^{\text{ML}(1)}[\vv], \dots, u^{\text{ML}(L)}[\vv]\right]$ and $k$ is once again the Gaussian kernel.

\subsection{Cross-validation}
All hyperparameters used in the model are estimated solely on the training set. The width $\gamma$ and spacing $\Delta$ hyperparameters for the
artificial Gaussians potential were optimized for all molecules at the same time, with the resulting optimal values being $\gamma=0.2~\angstrom$ and
$\Delta=0.08~\angstrom$, while the kernel width $\sigma$ and the regularization parameter
$\lambda$ were optimized individually for each molecule. In both cases the hyperparameter optimization was performed using cross-validation
\cite{hansen2013assessment}. After training, the model is fixed and is applied unchanged on the out-of-sample test set.

\subsection{Functional and density driven error}

In order to more accurately measure the error of ML-HK map, we can separate out
the effect of the error in the functional $F$ and the error in the density $n(\vr)$
from the resulting error in the total energy, as shown in~\cite{kim2013understanding}.
Let $\tilde{F}$ denote an approximation of the many body functional $F$,
and $\tilde{n}(\vr)$ the resulting approximate ground-state density
when $\tilde F$ is used in the Euler equation. Defining
$\tilde E[n]=\tilde F[n] + \int d^3r \, n(\vr) \, v(\vr)$ yields
\begin{equation}
  \Delta E = \tilde E[\tilde{n}]- E[n] = \Delta E_\mathrm{F} + \Delta E_\mathrm{D}
\end{equation}
where $\Delta E_\mathrm{F} = \tilde{F}[n] - F[n]$ is the functional-driven error,
and $\Delta E_\mathrm{D} = \tilde E[\tilde{n}] - \tilde{E}[n]$ is the density-driven error.
We will use these additional error definitions to measure the accuracy of the ML-HK map.

\subsection{Results}
Here we revisit the results of applying the ML-HK map to predict electron densities and energies for a series of small 3D molecules. 
Using a set of test molecules, we compare the predictions of the ML model with the KS-DFT results obtained as described in Section~\ref{sec:ks-dft}.
Additionally we compare the ML-HK map using the Fourier basis representation the approach of directly mapping from the potential grid
$v(\vr)$ to the total energy. We call this approach the ML-KS model, with the resulting KRR formulation being
\begin{align}
  E^{\text{ML}}[v] = \sum_{i=1}^M \alpha_i k(v, v_i).
\end{align}

The first and most basic molecular prototypes used to evaluate the model were $\mathsf{H_2}$ and $\mathsf{H_2O}$, with the datasets 
being generated with one and three degrees of freedom respectively. For more details on the composition of the datasets
see~\cite{brockherde2017bypassing}.

For the evaluation of the models, for both datasets a random sample of 50 molecules was
taken as an out-of-sample test set. Additionally, for each of the molecules multiple subsets of varying sizes $M$ were chosen
out of the rest of the samples as training sets for both models.

\begin{table*}[htb]
\centering
\setlength{\tabcolsep}{5.7pt}     
\setlength{\cmidrulekern}{0.5em} 
  \begin{tabular}{ll @{\hspace{5\tabcolsep}} cc cccc}
\noalign{\smallskip}
& & \multicolumn{2}{c }{ML-KS} & \multicolumn{4}{c }{ML-HK}\\
\cmidrule(lr){3-4}\cmidrule(lr){5-8}
\noalign{\smallskip}
\noalign{\smallskip}
& & \multicolumn{2}{c } {$\Delta E$}
& \multicolumn{2}{c }{$\Delta E$} & \multicolumn{2}{c } {$\Delta E_\mathrm{D}^\mathrm{ML}$} \\
\cmidrule(lr){3-4}\cmidrule(lr){5-6}\cmidrule(lr){7-8}
\noalign{\smallskip}
\noalign{\smallskip}
Molecule & $M$ & MAE & max & MAE & max &MAE & max \\
\cmidrule(lr){1-8}
\noalign{\smallskip}
\noalign{\smallskip}
& 5 & 1.3 & 4.3 & 0.70 & 2.9 & 0.18 & 0.54\\
$\mathrm{H_2}$ & 7 & 0.37 & 1.4 & 0.17 & 0.73 & 0.054 & 0.16\\
& 10 & 0.080 & 0.41 & 0.019 & 0.11 & 0.017 & 0.086 \\
\noalign{\smallskip}
\cmidrule(lr){1-8}
\noalign{\smallskip}
\multirow{6}{*}{$\mathrm{H_2O}$} & 5 & 1.4 & 5.0 & 1.1 & 4.9 & 0.056 & 0.17 \\
& 10 & 0.27 & 0.93 & 0.12 & 0.39 & 0.099 & 0.59 \\
& 15 & 0.12 & 0.47 & 0.043 & 0.25 & 0.029 & 0.14 \\
& 20 & 0.015 & 0.064 & 0.0091 & 0.060 & 0.011 & 0.058 \\
\noalign{\smallskip}
\hline
\end{tabular}
\caption{\label{tab:results_3d_H2}
\textbf{Prediction errors on $\mathsf{H_2}$ and $\mathsf{H_2O}$.} Shown for increasing number of training points $M$
  for the ML-KS and ML-HK approaches. In addition, the estimated density-driven contribution to the error for the
  ML-HK approach is given. In all cases except at maximum M, the energy error in the ML-HK map is largely is the energy map,
  not the density map. Energies are in kcal/mol.}
\end{table*}

Since the size of some training subsets is very small, careful selection of the samples is required in order to ensure that the subset covers the complete range of geometries.
This is achieved via K-means sampling, which selects the $M$ training points so that the samples are nearly equally spaced in
the geometry space (see~\cite{brockherde2017bypassing}).

The performance of the ML-KS map is evaluated by comparing predicted total energy $E^{\text{ML}}[v]$ that is mapped directly from 
the Gaussians potential with the calculated KS-DFT energies. For the ML-HK map the total energy $E^{\text{ML}}[n]$ is obtained
by mapping from the predicted density $n^{\text{ML}}[v]$, which itself is predicted by mapping from the potential to the ground-state
density in a three-dimensional Fourier basis representation (using 25 basis functions for each dimension, for a total of 125000 basis
coefficients).

Table~\ref{tab:results_3d_H2} shows the resulting performance of the models when trained on datasets of varying sizes and evaluated
using the out-of-sample test sets. The mean average error (MAE) of energy predicted using the ML-HK map is significantly
smaller than that of the ML-KS map,
indicating that learning the potential-to-density map and subsequently learning the density-to-energy functional is easier
then directly learning the potential-to-energy map, at least when using our representations for the potential and density.

For $\mathsf{H_2O}$, in order to achieve similar accuracies as for $\mathsf{H_2}$ we need a larger training set, which is
expected due to the increased degrees of freedom and complexity of the $\mathsf{H_2O}$ molecule. Considering that the
MAE of PBE energies relative to CCSD(T) calculations is 1.2 kcal/mol for the water dataset, the additional error introduced
by the predicted PBE energies using the ML-HK map is negligible.

For larger molecules, the number of degrees of freedom is much higher, making it difficult to randomly generate conformers
that span the full configurational space. Therefore, large datasets of conformers for benzene, ethane and malonaldehyde were generated
using classical force field molecular dynamics. The MD simulations were performed at 300~K, 350~K, and 400~K using
the General Amber Force Field (GAFF)\cite{wang2004development} in the PINY\_MD package\cite{tuckerman2000exploiting}.
The MD simulations generate a large and varied set of geometries which are then sub-sampled using the K-means approach 
to obtain 2,000 points which constitute the training set for each of the molecules.

\begin{figure*}[htb]
  \includegraphics[width=\textwidth]{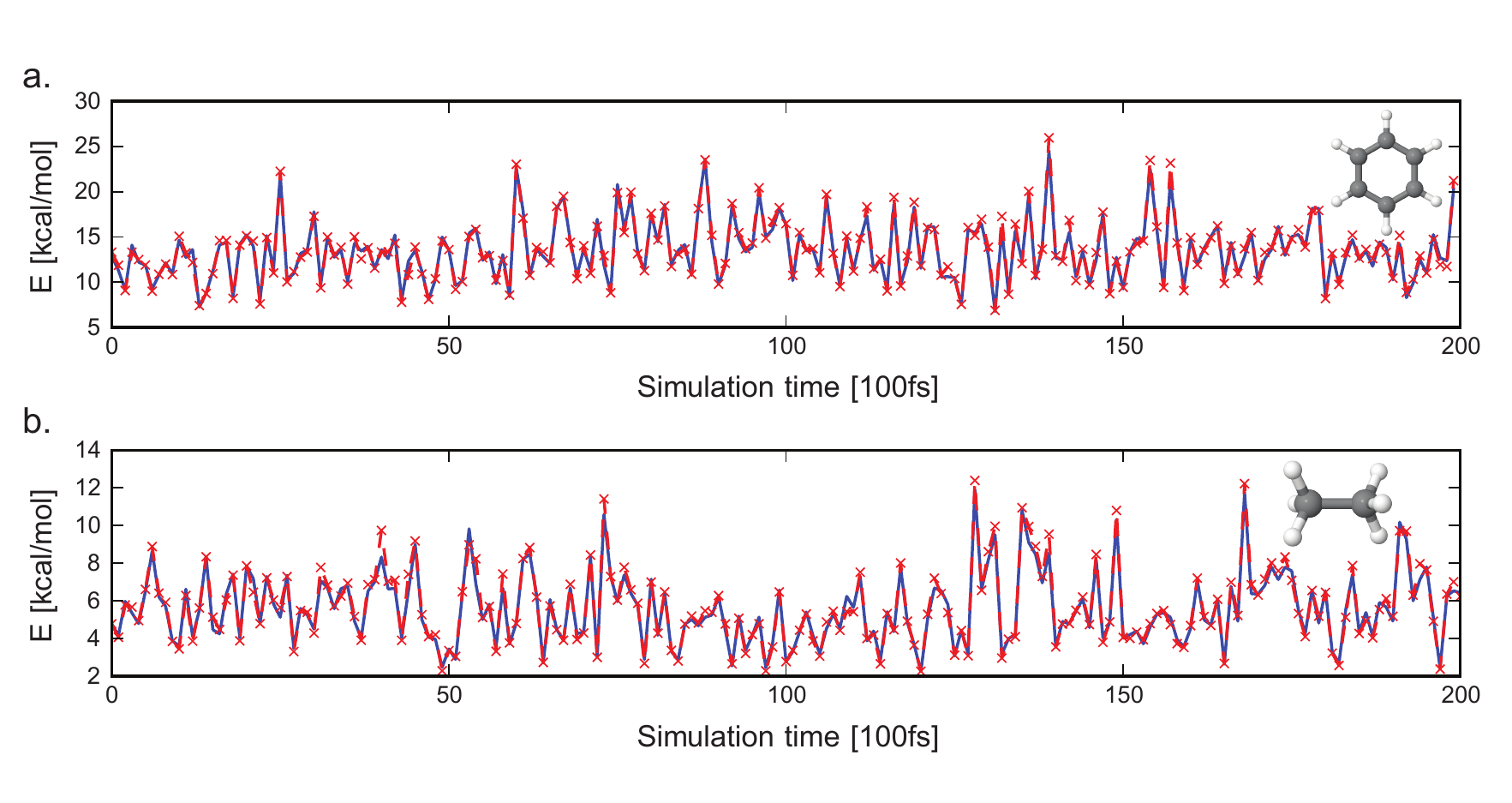}
  \caption{\label{fig:trajectories} \textbf{Energy errors of ML-HK along classical MD trajectories.} PBE values in blue, ML-HK values in red. 
  \textbf{a.} A 20~ps classical trajectory of benzene. \textbf{b.} A 20~ps classical trajectory of ethane. Figure adapted
  from Brockherde et al.~\cite{brockherde2017bypassing}.}
\end{figure*}

The training for the ML-HK and ML-KS models is performed as described above using reference KS-DFT calculations for the
density and total energy for a training set size of $M = 2000$ points. The performance of each models is then
tested on random samples taken from independent trajectories run at 300~K. Additionally, the performance is evaluated
for training sets composed of combined samples from MD trajectories at 300~K and a higher temperature, since this should increase
the span of the training set geometries, resulting in better generalization. The results of the evaluations for all three
molecules are shown on Table~\ref{tab:results_mol}.

\begin{table}
  \centering
  \setlength{\tabcolsep}{5.7pt}     
  \setlength{\cmidrulekern}{0.5em} 
  \begin{tabular}{ll @{\hspace{5\tabcolsep}} cccc}
    & \multirow{2}{*}[-0.6em]{\parbox{1.8cm}{Training trajectories}} & \multicolumn{2}{c}{$\Delta E$} & \multicolumn{2}{c}{$\Delta E_\mathrm{D}^{\mathrm{ML}}$}\\
    \noalign{\smallskip}
    \cmidrule(lr){3-4}\cmidrule(lr){5-6}
    \noalign{\smallskip}
    Molecule & & MAE & max & MAE & max\\
    \noalign{\smallskip}
    \cmidrule(lr){1-6}
    \noalign{\smallskip}
     & 300K & 0.42 & 1.7 & 0.32 & 1.5\\
    Benzene & 300K + 350K & 0.37 & 1.8 & 0.28 & 1.5\\
     & 300K + 400K & 0.47 & 2.3 & 0.30 & 1.8\\
    \noalign{\smallskip}
    \cmidrule(lr){1-6}
    \noalign{\smallskip}
     & 300K & 0.20 & 1.5 & 0.17 & 1.3\\
    Ethane & 300K + 350K & 0.23 & 1.4 & 0.19 & 1.1\\
     & 300K + 400K & 0.14 & 1.7 & 0.098 & 0.62\\
    \noalign{\smallskip}
    \cmidrule(lr){1-6}
    \noalign{\smallskip}
    Malonaldehyde & 300K + 350K & 0.27 & 1.2 & 0.21 & 0.74\\
    \noalign{\smallskip}
    \hline
    \noalign{\smallskip}
  \end{tabular}
\caption{\label{tab:results_mol}
\textbf{Energy and density-driven errors of the ML-HK approach on the MD datasets.} Errors are given in kcal/mol for different training trajectory combinations.
}
\end{table}

\begin{figure*}[htb]
  \includegraphics[width=\textwidth]{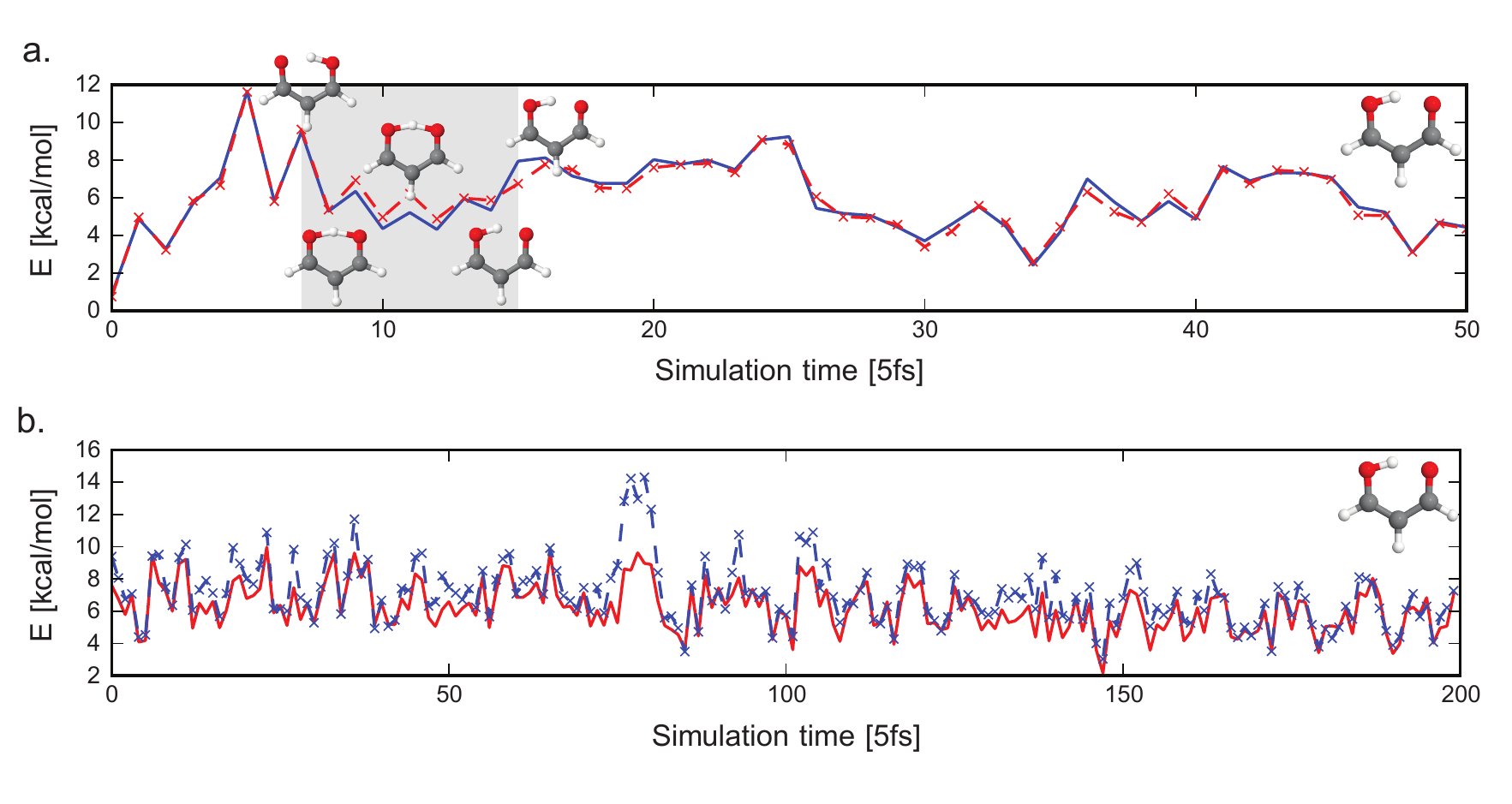}
  \caption{\label{fig:trajectories_malonaldehyde} \textbf{Energy errors of ML-HK along ab-initio MD and ML generated trajectories.}
  \textbf{a.} Energy errors of ML-HK along a 0.25 ps ab initio MD trajectory of malonaldehyde. PBE values in blue, ML-HK values in red. The ML model correctly predicts energies during a proton transfer in frames 7 to 15 without explicitly including these geometries in the training set.
  \textbf{b.} Energy errors of ML-HK along a 1 ps MD trajectory of malonaldehyde generated by the ML-HK model. ML-HK values in red, PBE values of trajectory snapshots in blue.}
  Figure adapted from Brockherde et al.~\cite{brockherde2017bypassing}.
\end{figure*}

The MAE achieved on the test snapshots using the ML-HK map trained on a set of 2000 samples is consistently under 1 kcal/mol for each of the 
molecules. Additionally, learning from a training set combining samples from trajectories run at higher temperatures improves the performance,
however, which higher temperature brings the largest benefits depends on the molecule itself. Figure~\ref{fig:trajectories} visualizes the
differences between the energies from the ML-HK map and the KS-DFT calculations on the test sets for benzene and ethane.
The versatility of the ML-HK map is further demonstrated by Figure~\ref{fig:trajectories_malonaldehyde}a, where the model manages to
interpolate the energies for a proton transfer in malonaldehyde, even though such geometries were not generated by the classical force field MD and thus were not present in the training set.

Finally, the ML-HK map for malonaldehyde can also be used to generate an MD trajectory at 300~K, obtained by finite-difference approach
to determine the atomic forces from the predicted energies (see Figure~\ref{fig:trajectories_malonaldehyde}b). Despite some molecular
configurations where the energy is underestimated (maximum absolute error of 5.7~kcal/mol), the resulting forces are still large enough
to bring the molecule to the equilibrium, resulting in a stable trajectory. Being able to run MD simulations using the ML-HK map can
greatly reduce the computation effort compared to running MD simulations with DFT, while at the same time providing results that are
very close to DFT accuracy. It is important to note that there are multiple ML models capable of providing even more accurate
MD simulations by directly learning the force field~\cite{chmiela2017machine,chmiela2018towards,grisafi2018symmetry,
glielmo2018efficient,zhang2018deep,christensen2018operators, schutt2017schnet}, however this experiment demonstrates that stable
MD simulations can also be produced using a force free machine learning model.

\section{Discussion}

The work revisited in this submission was one of the first ML models capable of efficiently and accurately
predicting both the electron density and total energy of small atomistic systems. While there are
many machine learning models that directly predict the total energy with greater accuracy,
predicting the density as an intermediate step opens up many different possibilities, such as
using the density as an universal descriptor to predict other
properties (due to the Hohenberg-Kohn theorem\cite{hohenberg1964inhomogeneous}) or
plugging the predicted density directly into DFT codes to perform calculations.

Here we showed once again in more detail how the problem of modelling a 3D electron density can
be simplified by using a basis representation and learning to predict the basis coefficients, and how for
orthogonal basis functions problem of predicting the coefficients can be decomposed into independent
learning problems for each of the coefficients. This greatly improves the simplicity and efficiency
of the resulting machine learning model.

Since the model is not dependent on the reference calculations used for training, the ML-HK map
can be trained using electron densities obtained by other methods besides KS-DFT, such as
Hartree-Fock or even CCSD densities. The same holds for the energy functional map, 
giving us the possibility of predicting total energies using electron densities obtained
using different levels of theory, potentially leading to even larger gains in computational
efficiency.

There is also plenty of room for improvement of the current model in the future. One of the main
challenges of the current model is its handling of the many symmetries in the data. While the potential
used as a descriptor is invariant to permutational symmetries, rotational symmetries have to be explicitly
tackled by rotating each of the molecules to optimally align with a reference geometry, which can often be
a source of noise, since the conformers often do not perfectly match the reference. Additionally, the
current descriptor has no way to handle mirror symmetries, meaning that the model would require a larger
number of training samples to learn these symmetries. The same holds for the Fourier basis representation,
which is also not invariant to symmetries of the O(3) group. Consequently, an obvious avenue for improvement
would be to incorporate descriptors that are invariant to the various symmetries, both for the atomistic systems~\cite{schutt2014represent,
de2016comparing, faber2018alchemical} and the electron densities~\cite{eickenberg2018solid}.

Finally, since the ML-HK map is independent from the subsequent density-to-energy map, the model can
be easily extended to predict various other properties of the atomistic system, for example by replacing
the density-to-energy map by a density-to-forces map, which has the potential of drastically improving
the accuracy the MD simulations produced using the machine learning model.

This demonstrates the flexibility of the formulation of the ML-HK map and provides many potential
directions in which the results discussed here can be further expanded and improved.

\section{Acknowledgments}
We thank IPAM at UCLA for repeated hospitality. Work at UC Irvine supported by NSF CHE-1464795. KRM and FB thank the Einstein Foundation 
for generously funding the ETERNAL project. This research was supported by Institute for Information \& Communications Technology Promotion
and funded by the Korea government (MSIT) (No. 2017-0-00451, No. 2017-0-01779). Work at NYU supported by the U.S.\ Army Research Office under 
contract/grant number W911NF-13-1-0387 (MET and LV). Ab initio trajectory was run using High Performance Computing resources at 
NYU\@. Other DFT simulations were run using High Performance Computing resources at MPI Halle.


\end{document}